\documentclass[preprint,12pt,3p]{elsarticle}
\usepackage{amsmath,amssymb,amsthm,graphicx,url}
\usepackage{mathtools} 

\usepackage{todonotes}
\usepackage{balance}
\usepackage{caption}
\usepackage{subcaption} 
\usepackage{booktabs}
\usepackage{multirow}
\usepackage{setspace}
\usepackage{float}
\usepackage{times}

\usepackage{setspace}

\usepackage{xspace}

\newcount\Comments  
\usepackage{color}
\newcommand{\kibitz}[2]{\ifnum\Comments=0\textcolor{#1}{#2}\fi}
\definecolor{amber}{rgb}{1, 0.49, 0}

\newcommand{\mybeta}{$\beta$ }

\definecolor{green}{rgb}{0,0.65,0.0}

\makeatletter
\def\ps@pprintTitle{%
   \let\@oddhead\@empty
   \let\@evenhead\@empty
   \def\@oddfoot{\reset@font\hfil\thepage\hfil}
   \let\@evenfoot\@oddfoot
}
\makeatother

\usepackage{amssymb}

\journal{Future Generation Computer Systems}

\begin{document}

\begin{frontmatter}

\title{Decentralized Optimization of Vehicle Route Planning -- A Cross-City Comparative Study}



\cortext[cor1]{Corresponding author}




\author[a]{Brionna Davis}
\author[a]{Grace Jennings}
\author[b]{Taylor Pothast}
\author[c]{Ilias Gerostathopoulos}
\author[d]{Evangelos Pournaras}
\author[e]{Raphael E. Stern\corref{cor1}}
\ead{rstern@umn.edu}

\address[a]{Department of Civil and Environmental Engineering, Vanderbilt University, 2301 Vanderbilt Place, Nashville, TN 37235, USA.}
\address[b]{Department of Human and Organizational Development, Vanderbilt University, 2301 Vanderbilt Place, Nashville, TN 37235, USA.}
\address[c]{Department of Computer Science, Vrije Universiteit Amsterdam, De Boelelaan 1105, 1081 HV Amsterdam, Netherlands}
\address[d]{School of Computing, University of Leeds, Leeds LS2 9JT, UK}
\address[e]{Department of Civil, Environmental, and Geo- Engineering, University of Minnesota, 500 Pillsbury Dr. SE, Minneapolis, MN 55455, USA.}


\begin{abstract}
New mobility concepts are at the forefront of research and innovation in smart cities. 
The introduction of connected and autonomous vehicles enables new possibilities in vehicle routing. Specifically, knowing the origin and destination of each agent in the network can allow for real-time routing of the vehicles to optimize network performance. However, this relies on individual vehicles being ``altruistic'' i.e., being willing to accept an alternative non-preferred route in order to achieve a network-level performance goal.
In this work, we conduct a study to compare different levels of agent altruism and the resulting effect on the network-level traffic performance. Specifically, this study compares the effects of different underlying urban structures on the overall network performance, and investigates which characteristics of the network make it possible to realize routing improvements using a decentralized optimization router.
The main finding is that, with increased vehicle altruism, it is possible to balance traffic flow among the links of the network. We show evidence that the decentralized optimization router is more effective with networks of high load while we study the influence of cities characteristics, in particular: networks with a higher number of nodes (intersections) or edges (roads) per unit area allow for more possible alternate routes, and thus higher potential to improve network performance.
\end{abstract}


\end{frontmatter}

\section{Introduction}


New mobility concepts are at the forefront of research and innovation in smart cities and are enabled by advances in intelligent infrastructures.
The shift toward an \textit{autonomous vehicle} (AV) fleet means that we will soon have the possibility to control the routes that individual vehicles take. 
Even before AVs are prevalent on our roadways, vehicle connectivity via smartphone apps (e.g., Waze, Google Maps, Apple Maps, Nokia HERE, etc.) already make it possible to suggest individualized routes for each driver in the network, and optimize these routes based on some desired network state. 
This is often referred to as a \textit{system optimal} (SO) route assignment~\cite{bell1997transportation}. 
However, compliance to these routes is not enforceable. 
Furthermore, the extent to which these companies coordinate to optimize at the system level is not known.

If routes are assigned to drivers based on what is best to reach the network state under some SO criteria, this may not be the route that is best for each individual driver. Instead, depending on how selfish they are, an individual driver may select the \textit{user equilibrium} (UE) route, i.e. the route that is optimal for each individual driver based on a greedy assessment of the route options~\cite{bell1997transportation}. While selfish drivers may select the route that is best for them, more altruistic drivers may be willing to accept the SO route assignment, while some drivers who have both selfish and altruistic traits may select some hybrid route that has elements of both the SO and the UE route.

Finding SO routes has been an area of significant research over the last several decades and generally is considered under \textit{dynamic traffic assignment} (DTA), which dates back to the late 1970s~\cite{merchant1978model, merchant1978optimality}. An excellent summary of early DTA efforts is provided by Peeta and Ziliakopoulos~\cite{peeta2001foundations}. Much of the existing early literature on DTA focuses on mathematical programming based solutions to routing vehicles in fixed time steps~\cite{merchant1978model}. Other approaches to finding optimal routing include the formulation as an optimal control problem~\cite{friesz1989dynamic, ran1993new, chen1998game}.


Many influential works have provided strategies for identifying optimal routing~\cite{mahmassani1993network, jahn2005system, groot2014toward, shen2014system}. More recently, the possibilities of conducting DTA with AVs has been considered, with many aspects of the problem having been considered~\cite{zhu2015linear, levin2015intersection, levin2016multiclass}. 
This work builds on prior efforts in this area by considering an adaptive routing framework called TRAPP which applies decentralized agent-based planning and optimization to optimize the operational efficiency of a city \cite{trapp}.
We hence utilize a socio-technical approach that goes beyond the traditional approach of speed limits and exploits cooperation between agents to maximize the utilization of the existing infrastructure.
Specifically, within this context, we consider the application of this adaptive routing framework for a fleet of autonomous vehicles, and consider the impact of each vehicle's selfishness (i.e., willingness to accept a non-optimal route to improve the goal of network optimality) on the overall performance. 




The overall idea behind our work is to improve vehicle routing on a network by exploring the trade off between optimizing global and local objectives that the agents consider when selecting a route. In this context, a global objective is a system-level objective such as reducing CO$_2$ emissions or balancing the load assigned to each link in the network, while a local objective is one that is specific to an individual vehicle, such as travel time. The general assumption is that in pursuing the global objective, the efficiency of the entire network is improved. In particular, we assume that even if \textit{some} of the agents pursue a mix of UE and SO routing, this can benefit all agents in the system, including the selfish ones, and still improve the efficiency of the network. 

The specific mechanisms to incentivize travel behavior or nudge drivers to become ``altruistic'' and take SO routes are the focus of several other works~\cite{kazhamiakin2015using, klein2018emergence, ringhand2018make}, and are beyond the scope of this work. Instead, assuming an incentivization mechanism exists, we try to answer the question: What degree of altruism is required by the agents to see system-level benefits and to what extent is the required altruism dependent on the city's urban structure and traffic level in the network?







In the context of employing multi-agent learning to optimize route planning, in this paper, we make the following contributions: 

\begin{itemize}
    \item We present a study that compares different altruism levels of agents (autonomous vehicles) and their effect on the overall traffic performance. 
    \item We compare such effects across cities with different characteristics and investigate which characteristics make it possible to observe positive traffic effects via increased altruism levels.
    \item The understanding of how different traffic levels in the network influence the effectiveness of alternative optimized routes. 
    \item The applicability of an open-source software framework\footnote{TRAPP, available at \url{https://github.com/iliasger/TRAPP}} to different large-scale urban transport networks. The framework integrates a well-known traffic simulator with a decentralized agent-based planner and allows for performing studies that employ multi-agent learning in optimizing route planning.
\end{itemize}



The remainder of the paper is outlined as follows. In Section~\ref{sec:tech_background}, technical background on the simulation tool, the optimization tool, and their integration are presented. The design of our study is presented in Section~\ref{sec:exp_design}; the study results and their interpretation are presented in Section~\ref{sec:exp_results}. Finally, the conclusions are provided in Section~\ref{sec:conclusions}, and possible directions for future work are also discussed.



\section{Technical Background}\label{sec:tech_background}

In order to investigate the effect of local and global objectives in dynamic traffic assignment via agent-based planning, we performed a simulation study. 
In this section, we describe in detail the traffic simulator--SUMO--and the agent-based planning framework--EPOS--we used and how they were integrated for the purposes of our study.  

\subsection{Traffic Simulation with SUMO and TraCI}

SUMO (Simulation of Urban MObility) is a well-known open-source microscopic traffic simulator~\cite{lopez_microscopic_2018}. 
It can be used for simulating up to hundreds of thousands of vehicles in complex, realistic city networks (which can be extracted e.g. from OpenStreetMaps) that include traffic signals, multiple lanes per street, speed limits per street, different types of streets (for cars, bicycles or pedestrians), among other features~\cite{7906642,8275627}. 
High realism is achieved in SUMO by simulating acceleration and deceleration of cars in traffic lights and intersections, vehicle manoeuvres, and by allowing for different driving styles (e.g. aggressive drivers).
All the above features make SUMO a mature tool for performing traffic simulation and analysis. 

In our experiments, we relied on a module of SUMO that can be used for controlling a SUMO simulation via Python~\cite{Wegener:2008:TIC:1400713.1400740}. 
TraCI (Traffic Control Interface) provides a convenient interface for both inspecting different attributes of the simulation and calculating performance indicators out of them (e.g. duration of each car trip, number of cars on each street at each simulated time point), but also for intervening and changing aspects of the SUMO simulation. 
For instance, TraCI allows to change the route assigned to a car--a feature that we used in our experiments. 

In order to experiment with dynamic route assignment, we implemented three different routers in Python. 
All routers rely on the internal representation of a city network as a graph and perform a Dijkstra algorithm to find the shortest path between an initial position A and a destination B. 
The difference of the routers rely on what they consider as cost of an edge (street). 
In the first router such a cost is the length of the street, resulting in routes of minimal overall length. 
In the second router, the cost is the inverse of the maximum speed allowed on a street, resulting in routes of maximum overall speed. 
In the third router, the cost is formed by the length of a street divided by the maximum speed allowed on the street, resulting in routes of minimal length \textit{and} maximum overall speed. 

In our experiments, each router produced a single route for each trip, hence each car could select among three different routes to navigate from A to B. 
Which route to choose was a decision that involved agent-based planning via EPOS, described next. 

\subsection{Traffic Optimization with EPOS}

EPOS (Economic Planning and Optimized Selections) is a decentralized multi-agent optimization framework written in Java~\cite{epos2018}. 
EPOS can be used for efficiently solving complex multi-objective combinatorial problems via participatory collective learning. 
In particular, EPOS assumes that a number of agents needs to coordinate their decisions in order to effectively use a shared medium such a power grid or a set of streets. 
Each agent's decisions may influence the decision of other agents, i.e. non-linear cost functions.
The problem that EPOS solves is to allow each agent to take decisions that considers both local and global objectives with the minimum amount of interaction with the other agents. This is achieved by having agents in EPOS self-organize in tree topologies over which they can perform efficient aggregation and decision-making in an iterative fashion: consecutive child-parent interactions in the bottom-up phase, followed by parent-child interactions in the top-down phase. 
In the following, we will describe EPOS only to extent necessary for this study, we refer the interested reader to \cite{epos2018} for more details. 

\begin{figure}
  \centering
  \includegraphics[width=15.2cm]{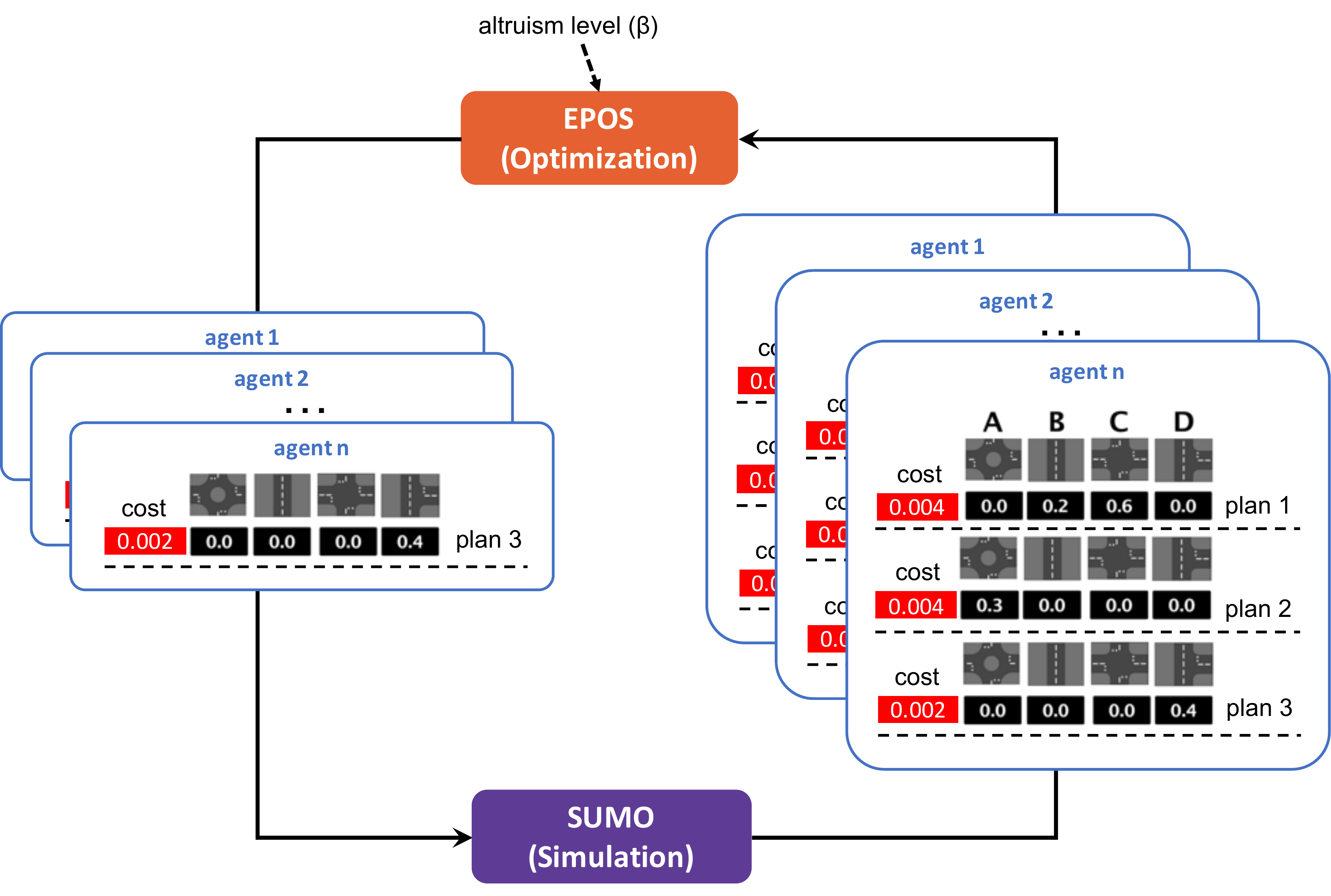}
  \caption{Plan specification and selection via inter-operation of SUMO and EPOS in our study.}
  \label{fig:trapp}
\end{figure}

In this study, an EPOS agent is a self-driving car.
Decision making in EPOS involves selecting a \textit{plan} from a finite set of plans for each agent. 
In our setting, a plan corresponds to a \textit{route} from a position A to a destination B. 
As explained in the previous subsection, we equipped each self-driving car with the ability to select among three possible routes, each corresponding to one of the three available routers (``minimum length'', ``maximum speed'', and ``combined length and speed'' router).
Nevertheless, our setting can be easily extended to accommodate more routes and routers, even routers that only serve specific cars (essentially creating agents that have more options).
EPOS is then used by the self-driving cars in our study so that each car selects one route to follow out of the three options they have (Figure~\ref{fig:trapp}). 

A plan is represented as a vector of real values in EPOS, each representing the ``contribution'' of the agent to the shared medium. 
In our study, the shared medium is the set of all streets in a city. 
A plan hence becomes a vector of real values containing the expected utilization of each street by the car in the corresponding route for a specific planning horizon (e.g. 30 minutes).
For instance, assuming a city consisting of only four streets, \texttt{A, B, C, D}, a route that only uses \texttt{A} is encoded as \texttt{X, 0, 0, 0}, where \texttt{X} depends on the expected occupancy of \texttt{A} (which is calculated based on the street length, vehicle length and expected time spent on the street). 
Plan 2 of agent n in Figure~\ref{fig:trapp} is a concrete such example with \texttt{X = 0.3}. 

Each plan comes with a a cost denoting the dislike of the agent for this plan. 
Agents can express their preferences for plans that they generate by lowering the cost of preferred plans. 
For instance, plan 3 for agent n in Figure~\ref{fig:trapp} is preferred over plans 1 and 2. 
In our study, we assign costs to plans based on agents' preferences mined from historical (simulated) runs of the self-driving cars. 
We describe this in detail in Section~\ref{sec:exp_design}.

In EPOS, an agent makes a plan selection based on three criteria: (i) global cost (GC), (ii) local cost (LC), and (iii) unfairness (U).
For our study, we model global cost as the variance of street utilization when considering the utilization contributed by all cars. 
(global cost is computed by EPOS by first summing up element-wise all the selected plans and then calculating the variance of the resulting vector.)
Our model of global cost is a natural one, since global cost represents what needs to be optimized at a system level.
In our case, this is the variance of street utilization, since we would like to balance the self-driving cars in the available streets in order to avoid traffic externalities.
Local cost is simply the cost of each plan, as provided by each agent (see Figure~\ref{fig:trapp}).
Finally, unfairness captures how important is for the agent that the overall solution creates equal dissatisfaction among the agents; it is computed as the variance over the cost of the selected plans.
The final cost of a plan is a weighted sum of the three criteria:
\begin{equation}\label{eq1}
    (1-\alpha-\beta)GC + \alpha U + \beta LC
\end{equation}
where $\alpha$ and $\beta$ are real values in [0,1].

Since the objective of each agent is to select the plan with the lowest final cost, it becomes apparent that setting $\alpha=0$ assumes that fairness concerns play no role in the optimization process.
In our study, we have always used $\alpha=0$ and experimented with different values of $\beta$.
It is important to note that high values of \mybeta represent more selfish agents, since they care more about their local cost than the global one. 
Conversely, lower values of \mybeta represent more altruistic agents.
On the extreme cases, $\beta~=~0$ models completely altruistic agents, whereas $\beta~=~1$ completely selfish ones.

\subsection{Integration of SUMO and EPOS}
In this study, we couple each car present in SUMO to an agent present in EPOS. 
Using the TRAPP framework~\cite{trapp} developed in our earlier work, we are able to run SUMO simulations which involve invoking EPOS at predefined time points (e.g. at the beginning of the simulation and also periodically with a predefined period). Every time EPOS is invoked, the Python-controlled simulation pauses and waits for the EPOS run to complete. 
After completion, the routes selected by EPOS are applied to the active simulation and the simulation resumes.

From the user's perspective, the simulation can be configured with different maps and different number of cars, and different values of the \mybeta parameter that controls the level of altruism of agents in EPOS. 
The duration of the simulation can also be configured, along with data to be logged (e.g. duration of trips, utilization of streets). 
In our experiments, in order to evaluate the effect of balancing the cars in a city network on the duration of trips---the main performance indicator for the passengers of self-driving cars---we log the durations of all completed trips.






\section{Study Design}\label{sec:exp_design}

Out of the many trade-offs in decentralized optimization of traffic, in this paper we focus on the trade-off between optimizing for local costs of the individual agents and the global costs of the network as a whole to answer the questions: 

\begin{enumerate}
\item Can increasing the altruism of traffic agents lead to positive traffic effects related to reduced trip times? 
\item If yes, what is the level of altruism necessary for observing such positive effects under varied level of traffic in the network?
\item Which are the city attributes and characteristics that determine whether such positive effects are observed?
\end{enumerate}


To investigate the above questions, we used the TRAPP framework that integrates SUMO and traCI, a well-known traffic simulator, with EPOS, a decentralized optimization framework, as described in the previous section. 
We performed several simulation runs using different values of the \mybeta parameter of EPOS on different traffic settings-- city maps and traffic levels-- as described in the rest of this Section. 
We also provide the open-source software TRAPP, together with concrete replication instructions online\footnote{\url{https://github.com/iliasger/TRAPP/blob/experiments/Beta_Alpha_Testing_README.md}}.


As explained in the previous Section, changing \mybeta leads to changing the altruism level of EPOS agents.
For each \textit{traffic setting} (explained next), we performed a systematic parameter sweep of $\beta$ starting from 1 (corresponding to completely selfish agents) to 0 (corresponding to completely altruistic agents) with a step of 0.1.
We kept all other TRAPP parameters constant in order to observe the effect of changing \mybeta alone.
For each \mybeta value, we performed 5 simulation runs with different random seeds, which affect the initial positions and destinations of cars, to obtain statistical validity. 
Hence, for each traffic setting we performed a total of $5x11=55$ simulation runs. 


We consider a \textit{simulation run} to be the simulation of a specific number of cars on the specific city network for a set amount of time--the \textit{simulation horizon}. 
The initial positions of cars are selected based on the population distribution of city districts; their destinations are randomly selected. 
EPOS is invoked only at the beginning of each run, with a planning horizon equal to the simulation horizon, and selects one route for each car.
Then, cars follow their routes without further planning. 
Once a car completes its trip, it picks another random destination and retrieves a route to it via its preferred router (the one with the lowest cost).
This ensures that the number of cars remains constant for the whole duration of a run. 
After a run is completed, the durations of the fist trip of each car that was completed are analyzed to determine the effect of EPOS optimization with the particular \mybeta value on traffic.
In particular, we compute the average of all logged trip overheads, where a trip overhead is an actual trip duration divided by the theoretical trip duration if the car would drive alone in maximum speed at all times. 

\begin{table}[]
\begin{tabular*}{\textwidth}{l | c c c c c}
\toprule
City, State    & Size & Density & Total pop. & \# Zip codes &  \% Commute by car \\
\midrule
Manhattan, NY   & 51 km$^2$ & 19,658 p/km$^2$ & 1,002,576          & 31          & 5.8                   \\
Duluth, MN   & 236 km$^2$ & 493 p/km$^2$ & 116,688           & 11          & 74.4                   \\
Annapolis, MD   & 21 km$^2$ & 2,945 p/km$^2$ & 61,837            & 4           & 73                \\
Boulder, CO    & 540 km$^2$ & 224 p/km$^2$ & 120,932           & 6           & 64.3           \\
\bottomrule
\end{tabular*}
\caption{Summary statistics for the four cities compared in our study.}
    \label{tab:t1}
\end{table}


A \textit{traffic setting} in our study is defined by the city map and the number of cars in the simulation. 
To understand the impact of urban structure on the ability to optimize trip routing via EPOS, the study considers four cities, Annapolis, Boulder, Duluth, and the borough of Manhattan in New York City. All of the cities used for the comparative study are located in the United States for the consistency of ZIP codes, census data, and commute data. These cities were also chosen for their diversity of urban infrastructure, including their area, population size, population density, and street organization~(Table~\ref{tab:t1}).  


The number of cars used for each simulation run was determined in the following way. 
We aim to compare cities based on the morning commute time. 
Hence, we first set the simulation horizon of a simulation run to 30 minutes, which roughly correlates to the average commute time in the US of 25.5 minutes.
Then, we calculate the total number of commuting trips using the percentage of drivers in each city that drive alone and the total population of the city from the 2010 census~(Table~\ref{tab:t1}).
This total number of commuting trips is divided by six, as typical morning peak traffic is from 6:30 am to 9:30 am, covering six half-hour time periods. Thus, the assumption that commuters are uniformly distributed in time over the morning commute is implicitly made. 
The resulting number of cars per city is depicted in Table~\ref{tab:t2}.
The number of cars and the simulation horizon thus represent a period of peak traffic of a city's morning commute.

\begin{table}[]
\centering
\begin{tabular*}{\textwidth}{l | c c c c}
\toprule
City, State & Nodes & Edges & Num. of cars & Density (number of cars / total street length)\\
\midrule
Manhattan, NY & 6,758 & 12,658 & 10,000 & 0.0093\\
Duluth, MN    & 5,607 & 14,317 & 14,000 & 0.0065\\
Annapolis, MD & 2,555 & 5,691  & 8,000 & 0.0147\\
Boulder, CO   & 7,740 & 16,625 & 13,000 & 0.0104\\
\bottomrule
\end{tabular*}
\caption{SUMO traffic settings for the four cities compared in our study.}
    \label{tab:t2}
\end{table}

\begin{figure}
  \centering
  \begin{subfigure}{6.5cm}
    \centering
    \includegraphics[width=6.5cm]{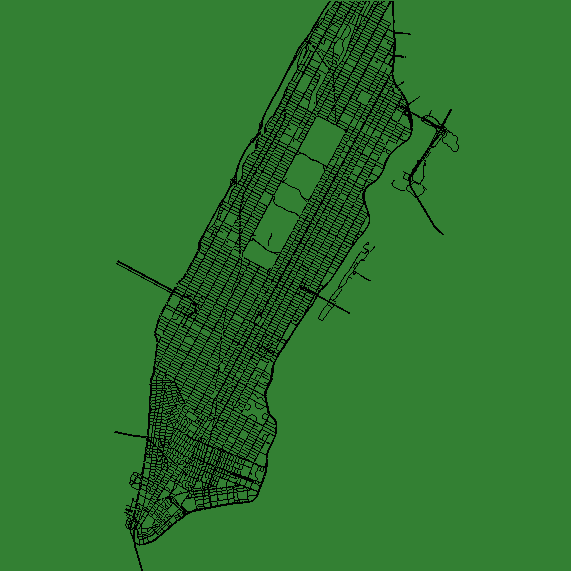}
    \caption{Manhattan}
    \vspace{2em}
    \label{fig:overhead1}
  \end{subfigure}
  \begin{subfigure}{6.5cm}
    \centering
    \includegraphics[width=6.5cm]{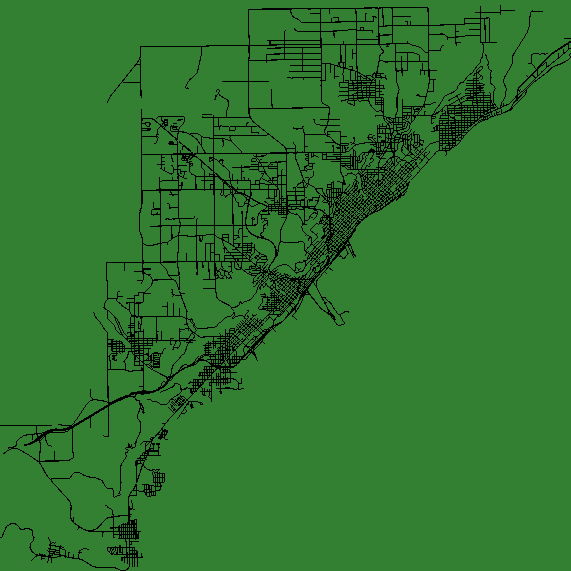}
    \caption{Duluth}
    \vspace{2em}
    \label{fig:overhead2}
  \end{subfigure}
  \begin{subfigure}{6.5cm}
    \centering
    \includegraphics[width=6.5cm]{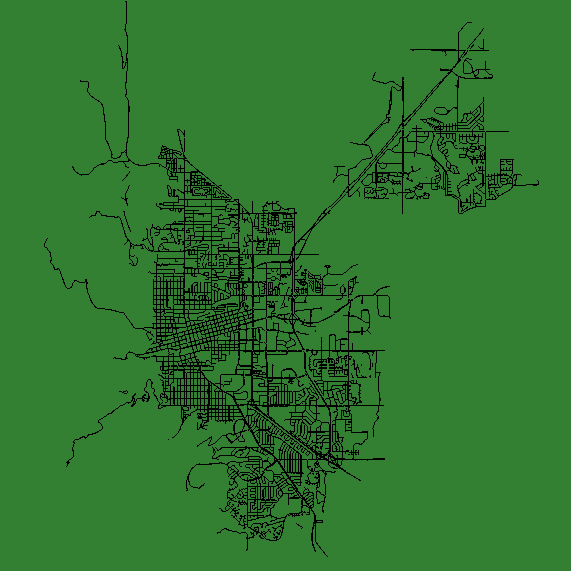}
    \caption{Boulder}
    \label{fig:overhead3}
  \end{subfigure}
  \begin{subfigure}{6.5cm}
    \centering
    \includegraphics[width=6.5cm]{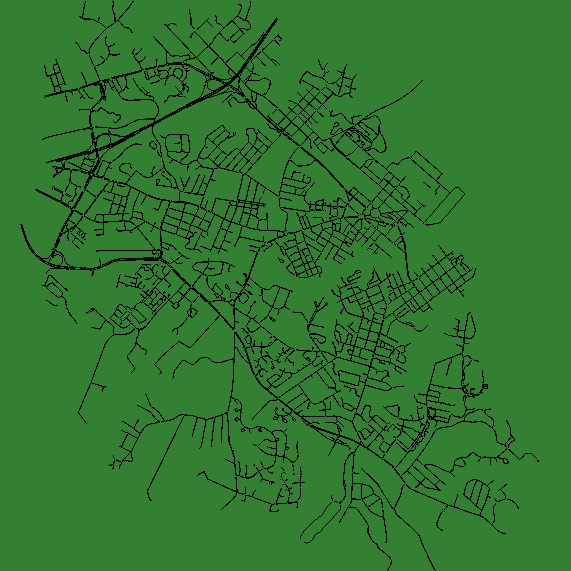}
    \caption{Annapolis}
    \label{fig:overhead4}
  \end{subfigure}
  \caption{The different maps used in our study.}
  \label{fig:maps}
\end{figure}

To make the study more realistic, it is important to consider additional factors that contribute to traffic flow, like traffic route patterns. We used a simplified model for traffic patterns
based on the assumption that during peak morning traffic people are driving from their places of residence to their places of work.
In our model, cities are divided into districts (squares in the map) with certain number of residents---population.
We computed a district's population by adding up the population of all ZIP codes whose centroids lie within the district. 
(Population per ZIP code is also obtained by the US 2010 Census data.)
Then, we calculated the distribution of a city's population in districts.
Using this distribution, cars' initial positions are assigned to districts; their specific position (a street within a district) is selected uniformly at random. 
Each trip destination is randomly selected within the city using a uniform distribution in space (without considering districts). This simplified destination selection method is used for lack of consistent data for the distribution of workplaces in cities. 

In order to prepare the SUMO maps of the different cities for our study we first used \textit{OSM Web Wizard} to extract SUMO networks from \textit{Open Street Maps}. 
We then cleaned each map using SUMO's \textit{NETEDIT} utility, a graphical network editor, to delete any unnecessary edges that extended beyond the each city's ZIP codes. Finally, we used SUMO's \textit{NETCONVERT} utility to remove any edges that could not accommodate passenger vehicles, removing any edges for buses, bikes, pedestrians, and trains from the maps.
The number of nodes (intersections) and edges (streets), along with a custom density metric of the resulting SUMO maps are listed in Table~\ref{tab:t2}, while the different maps are depicted in Figure~\ref{fig:maps}.

Finally, for each traffic setting, we conducted a number of baseline simulation runs to mine the cost that each agent associates with each router. 
The difference from the normal simulation runs that were used to derive the results is that in the baseline runs EPOS was not invoked. 
Instead, each car was selecting a router at random to perform a trip and logging the trip's overhead. 
For each traffic setting, we performed 100 baseline runs with different random seeds. 
We then calculated the average trip overhead per car per router, which became the \textit{cost} of that particular router for that car. 
These costs were used to characterize the local objective of each agent in EPOS.

\section{Study Results}\label{sec:exp_results}
In this section, we summarize the results obtained from running TRAPP on four different cities with differing urban structure. As already mentioned, for each traffic setting (city map and number of cars), 5 simulation runs were conducted for each of the 11 selfishness ($\beta$) values, ranging from $\beta~=~0$ to $\beta~=~1$. The resulting traffic performance in each traffic setting and \mybeta value is described in this section.

\begin{table}[t]
\centering
\begin{tabular}{l | c c c}
\toprule
Metric & Minimum & Maximum & Range (Maximum - Minimum) \\
\midrule
local cost    & 0.07438 & 0.20539 & 0.13100\\
global cost   & 0.00013 & 0.00131 & 0.00118\\
trip overhead & 2.67147 & 6.29150 & 3.62002\\
\bottomrule
\end{tabular}
\caption{Cross-city statistics of the three metrics in our study.}
    \label{tab:t3}
\end{table}

Note that for each of the result metrics (local cost, global cost, and trip overhead) we present normalized values to facilitate cross-city comparisons and interpretation. 
We perform a min-max normalization in which we first subtract the minimum value (observed for the metric across all cities) from the value being transformed and then divide by the value range of the metric (obtained by subtracting the minimum value observed for the metric across all cities from the maximum one). 
For reference, Table~\ref{tab:t3} provides an overview of the minimum, maximum, and corresponding ranges of the three metrics. 

\begin{figure}[!h]
  \centering
  \begin{subfigure}{6.5cm}
    \centering
    \includegraphics[width=6.5cm]{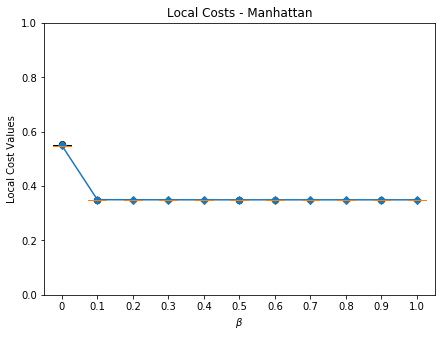}
    \caption{Manhattan}
    \vspace{2em}
    \label{fig:localcost1}
  \end{subfigure}
  \begin{subfigure}{6.5cm}
    \centering
    \includegraphics[width=6.5cm]{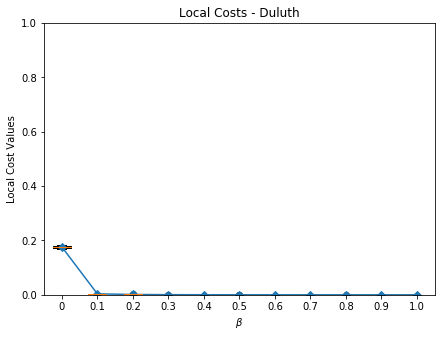}
    \caption{Duluth}
    \vspace{2em}
    \label{fig:localcost2}
  \end{subfigure}
  \begin{subfigure}{6.5cm}
    \centering
    \includegraphics[width=6.5cm]{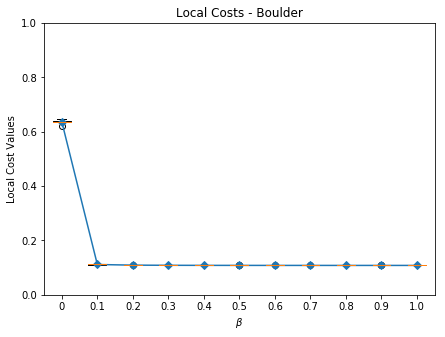}
    \caption{Boulder}
    \label{fig:lo calcost3}
  \end{subfigure}
  \begin{subfigure}{6.5cm}
    \centering
    \includegraphics[width=6.5cm]{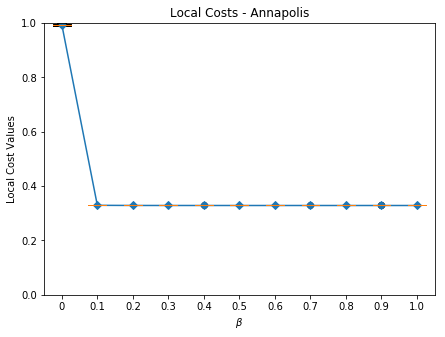}
    \caption{Annapolis}
    \label{fig:localcost4}
  \end{subfigure}
  \caption{Boxplots of local cost for all four cities as a function of selfishness $\beta$. Blue dots represent averages (over 5 runs). The trend shows that in all four cities, the local cost is constant except for the completely altruistic case ($\beta = 0$), where local cost is significantly increased.}
  \label{fig:localcost}
\end{figure}

\subsection{Local cost}
The \textit{local cost} of a run represents the average of the costs of the selected routes for the run, as reported by EPOS.
This gives an estimate of how much agents are dissatisfied --- lower values of local cost are better. 
Figure~\ref{fig:localcost} depicts the (normalized) local costs of all runs performed for each traffic setting and for each \mybeta value.

As a first observation, local cost is not the same for all cities, which captures the fact that some settings are inherently more costly to navigate than others. 
This is of course related also to the way we calculate the cost of a route for each car: each route inherits the cost of its router, which is calculated as the normalized average of trip overheads measured over all baseline runs that used the particular router.
Some settings yield inherently higher overheads and hence higher local costs than others. 
This is why local costs in Manhattan and Annapolis are higher and in Duluth and Boulder. 

The main pattern that emerges though for all cities is the following. 
As \mybeta is reduced, local cost remains almost unaffected until \mybeta reaches $0$.
At this point, local cost increases sharply.
The percentage increase between the highest value ($1$) and lowest value ($0$) of \mybeta is ~20\% for Manhattan, ~30\% for Duluth, ~77\% for Boulder, and ~66\% for Annapolis.

\begin{figure}[!t]
  \centering
  \begin{subfigure}{6.5cm}
    \centering
    \includegraphics[width=6.5cm]{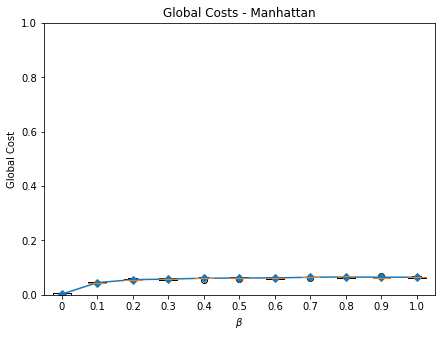}
    \caption{Manhattan}
    \vspace{2em}
    \label{fig:globalcost1}
  \end{subfigure}
  \begin{subfigure}{6.5cm}
    \centering
    \includegraphics[width=6.5cm]{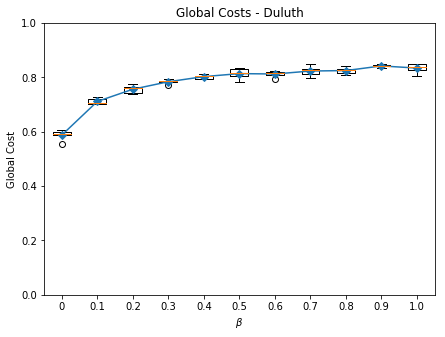}
    \caption{Duluth}
    \vspace{2em}
    \label{fig:globalcost2}
  \end{subfigure}
  \begin{subfigure}{6.5cm}
    \centering
    \includegraphics[width=6.5cm]{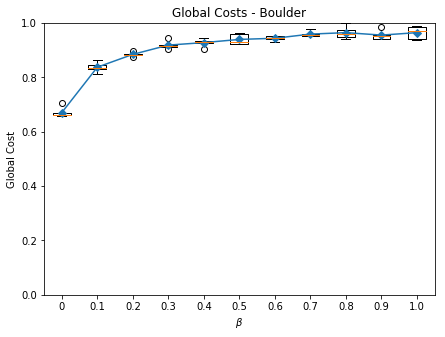}
    \caption{Boulder}
    \label{fig:globalcost3}
  \end{subfigure}
  \begin{subfigure}{6.5cm}
    \centering
    \includegraphics[width=6.5cm]{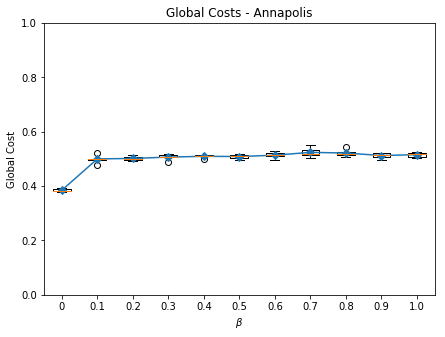}
    \caption{Annapolis}
    \label{fig:globalcost4}
  \end{subfigure}
  \caption{Boxplots of global cost for all four cities as a function of selfishness $\beta$, showing in increasing global cost with increased selfishness, meaning that when agents are more selfish, the overall cost for all agents is increased. Blue dots represent averages (over 5 runs).}
  \label{fig:globalcost}
\end{figure}

\subsection{Global cost}
The \textit{global cost} of a run represents the expected average variance of street utilization for the run, as reported by EPOS.
This gives an estimate of an important system-level objective of EPOS, i.e. to balance the presence of cars in the streets as much as possible (with the aim to avoid traffic externalities). 
Figure~\ref{fig:globalcost} depicts the (normalized) global costs of all runs performed for each traffic setting and for each \mybeta value.

Similar to local cost, global cost primarily depends on the setting, with Boulder having the highest global costs and Manhattan the lowest overall. 
With varying \mybeta, we see the inverse trend than local costs: global cost uniformly decreases going from \mybeta$=1$ to \mybeta$=0$. 
This is expected, since the more important the global objective becomes for EPOS, the more EPOS tries to decrease the variance in street utilization. 
However, in some settings the transition is more smooth than others (e.g. Duluth), where we observe sharper changes in the global cost (e.g. Manhattan). 
The percentage decrease between the highest value ($1$) and lowest value ($0$) of \mybeta is ~50\% for Manhattan, ~33\% for Duluth, ~34\% for Boulder, and ~20\% for Annapolis.

\begin{figure}[!h]
  \centering
  \begin{subfigure}{6.5cm}
    \centering
    \includegraphics[width=6.5cm]{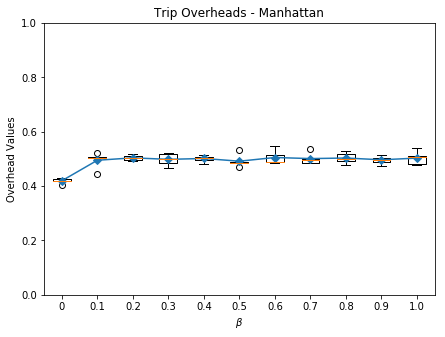}
    \caption{Manhattan}
    \vspace{2em}
    \label{fig:overhead1}
  \end{subfigure}
  \begin{subfigure}{6.5cm}
    \centering
    \includegraphics[width=6.5cm]{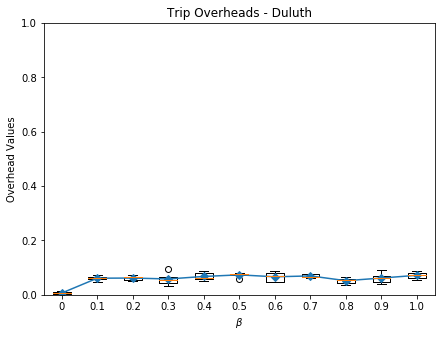}
    \caption{Duluth}
    \vspace{2em}
    \label{fig:overhead2}
  \end{subfigure}
  \begin{subfigure}{6.5cm}
    \centering
    \includegraphics[width=6.5cm]{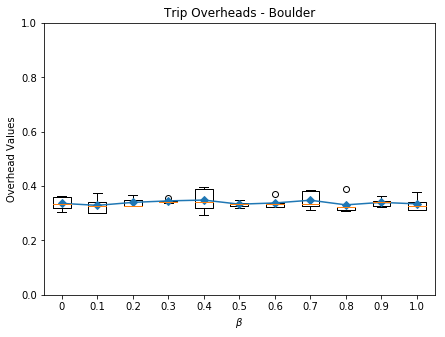}
    \caption{Boulder}
    \label{fig:overhead3}
  \end{subfigure}
  \begin{subfigure}{6.5cm}
    \centering
    \includegraphics[width=6.5cm]{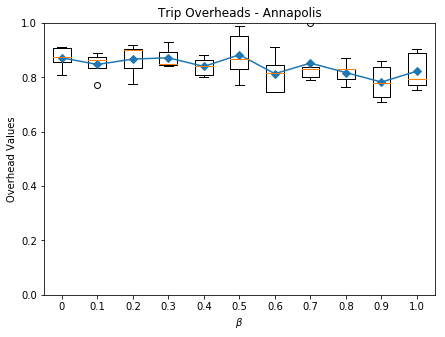}
    \caption{Annapolis}
    \label{fig:overhead4}
  \end{subfigure}
  \caption{Boxplots of trip overheads for all four cities showing an increasing trend with selfishness $\beta$ for Manhattan and Duluth (top), and no discernible trend with respect to selfishness $\beta$ for Boulder and Annapolis (bottom). Blue dots represent averages (over 5 runs).}
  \label{fig:overhead}
\end{figure}

\subsection{Trip overheads}

Trip overhead is the main metric we used to evaluate the effectiveness of our overall approach. Recall that trip overhead is computed as the actual trip duration divided by the theoretical trip duration if the car would drive at the maximum allowable speed throughout.
In particular, for each setting we calculated the mean of the trip overheads corresponding to the first trip of each car. 
This provides an estimate of the overall utility of the system -- with lower mean trip overhead corresponding to faster trips and hence higher system utility. 
Figure~\ref{fig:overhead} depicts the (normalized) mean trip overheads of all runs performed for each traffic setting and for each \mybeta value.

The first observation is that trip overhead clearly depends on the setting, with Annapolis having the highest values and Duluth the lowest. 
With varying \mybeta values, trip overhead shows no discernible trend, except for two cases: Manhattan with \mybeta$=0$ and Duluth with \mybeta$=0$ both show a statistically significant decrease in the trip overhead, as can be observed in Figures~\ref{fig:overhead1} and \ref{fig:overhead2}, respectively.
This trend is however not observed for Boulder (Figure~\ref{fig:overhead3}) and Annapolis (Figure~\ref{fig:overhead4}). 

\subsection{Influence of traffic level}

To clarify the influence of the route optimization on the trip overhead, it is hypothesized here that the effect of optimization can be easier observed when the network load, i.e. traffic level, is at a critical state. In other words, we assume that balancing of traffic flows decreases trip overhead only if there is a certain amount of traffic congestion in the network. Intuitively, alternative routes in a network free of traffic congestion are likely to increase travel times, while alternative routes in a congested networks are likely to reduce them. 

\begin{figure}[!h]
  \centering
  \begin{subfigure}{8cm}
    \centering
    \includegraphics[width=8cm]{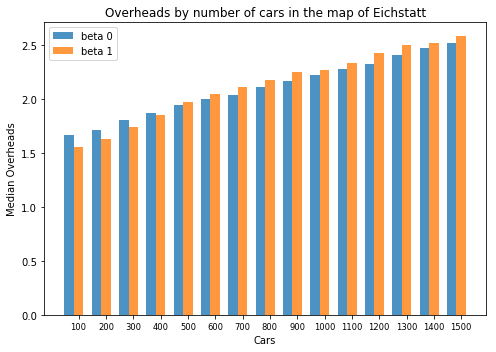}
    \caption{Trip overhead.}
    \vspace{2em}
    \label{fig:overheadsCrowdNav}
  \end{subfigure}
  \begin{subfigure}{8cm}
    \centering
    \includegraphics[width=8cm]{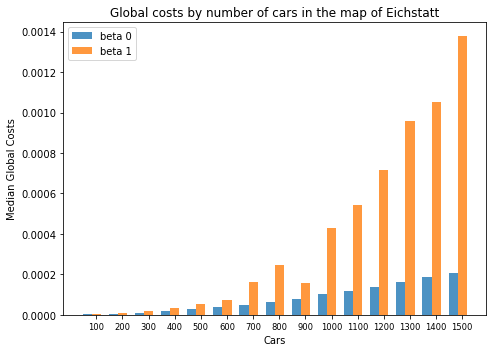}
    \caption{Global cost.}
    \vspace{2em}
    \label{fig:globalCostsCrowdNav}
  \end{subfigure}
  \begin{subfigure}{8cm}
    \centering
    \includegraphics[width=8cm]{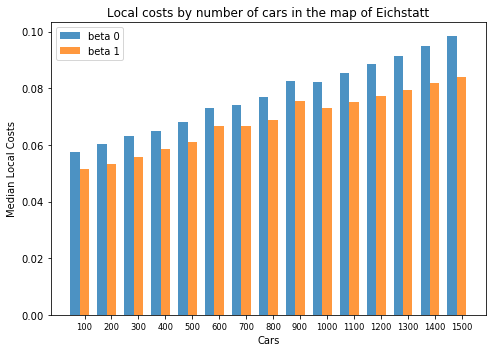}
    \caption{Local cost.}
    \label{fig:localCostsCrowdNav}
  \end{subfigure}
  \caption{Median results (over 5 random runs) for different number of cars in Eichst{\"a}tt. 
  }
  \label{fig:scalability}
\end{figure}

Given the large spectrum of the different performed experiments on large-scale networks, simulations are very computationally costly. For this reason, the aforementioned hypothesis is assessed on a smaller scale network. 
In particular, we use the city of Eichst{\"a}tt and varying number of cars ranging from 100 to 1500 with a step of 100 (see Figure~\ref{fig:scalability}).
Cars have random starting points and random destinations. 
Similar to the other experiments, when a car reaches a destination it picks another one to drive to; this ensures that the total number of cars in a run remains constant.  
Each run has a duration of 800 SUMO ticks, which is long enough to ensure that almost all the initial trips of each car are completed. 
Similar to the other experiments, for each run, we calculate the mean of the trip overheads of these trips, along with the local and global costs reported by EPOS.
EPOS is invoked at the beginning of each run.
In these experiments, we investigate the influence of total altruism ($\beta$=0) versus total selfishness ($\beta$=1, the baseline) under different number of cars.
Each setting (corresponding to number of cars and the beta value), is run five times with different random seeds.

Figure~\ref{fig:scalability} illustrates the results of median overhead, median global cost, and median local cost over the five runs per case. 
With respect to trip overheads, there is a critical state between 400 and 500 cars after which altruistic agents following the alternative routes consistently reduce the trip overhead. 
In a similar vein, we observe a sharp increase in the difference of global cost between altruistic and selfish agents for number of cars starting from 700 on. In other words, the reduction of trip overhead at a critical number of cars in the network is also associated with the optimization capacity of the traffic flows, indicated by the significant performance divergence between altruistic and selfish agents as the number of cars increases. 
On the contrary, local costs show a consistent difference between altruistic and selfish agents across all number of cars. 


\subsection{Interpretation of results}



In this subsection, we provide our interpretation of the results described in the previous subsections and attempt to answer the questions stated in our study design. 

First, the effect of changing the level of drivers' altruism (\mybeta value) is both clear and consistent across city settings for both the local and global cost. 
Local cost is practically unaffected for \mybeta values other than 0 and increases sharply when complete altruism is in place (\mybeta= 0).
In complete altruism, optimization in EPOS takes into account only the global objective (``reduce the variance of street utilization'') without taking the agents' preferences into account. 
Even slight consideration of agents' preferences (e.g. \mybeta = 0.1 or 0.2) drastically reduces the cost that the agents pay for the optimization to take place. 
The same ``all or nothing'' pattern is present in the evolution of global costs: even slight introduction of selfish behavior is enough to increase the global cost considerably. 
Still, in contrast to local costs, the global costs show a more gradual value change by increasing the altruism level. 

Looking at the results on trip overheads, we conclude that it is possible to use EPOS with altruistic agents, distribute the cars more evenly in the streets and, as a result, reduce the overall trip overheads, especially when the network is at a critical high traffic flow state. 

We clearly see such a positive traffic effect on average overhead values when setting \mybeta= 0 for Manhattan and Duluth. 
However, such an effect (i) is only present for the case of complete altruism (\mybeta= 0), and (ii) is not present in Boulder and Annapolis. 


\begin{table}[t]
\centering
\begin{tabular*}{\textwidth}{l | c c c c c}
\toprule
\multirow{2}{*}{City, State} & AVG  & Max & Min & Edges / & Nodes / \\
& street size & street size & street size & total street length & total street length\\
\midrule
Manhattan, NY & 84 m  & 2777 m & 0.1 m & 0.0117 & 0.0062\\
Duluth, MN    & 149 m & 5460 m & 0.1 m & 0.0066 & 0.0026\\
Annapolis, MD & 95 m  & 2764 m & 0.1 m & 0.0104 & 0.0047\\
Boulder, CO   & 74 m  & 4424 m & 0.1 m & 0.0133 & 0.0062\\
\bottomrule
\end{tabular*}
\caption{Further characteristics of the traffic networks of the cities considered in our study.}
    \label{tab:t4}
\end{table}

While assessing the critical traffic flow state of these urban networks is computationally challenging and out of the scope of this paper, we turn our attention to the characteristics of different city maps.
In particular, we calculated the average, maximum, and minimum street size, and also observed the whole distributions of the 
street sizes per city (not shown here).
We also divided the number of edges (streets) and nodes (intersections) by the total street length to obtain metrics of density of the map. 
The different measurements are depicted in Table~\ref{tab:t4} and there is no correlation between the values of these characteristics and the fact that balancing is effective in Manhattan and Duluth and not in Annapolis and Boulder. 

\begin{figure}
  \centering
  \begin{subfigure}{8cm}
    \centering
    \includegraphics[width=8cm]{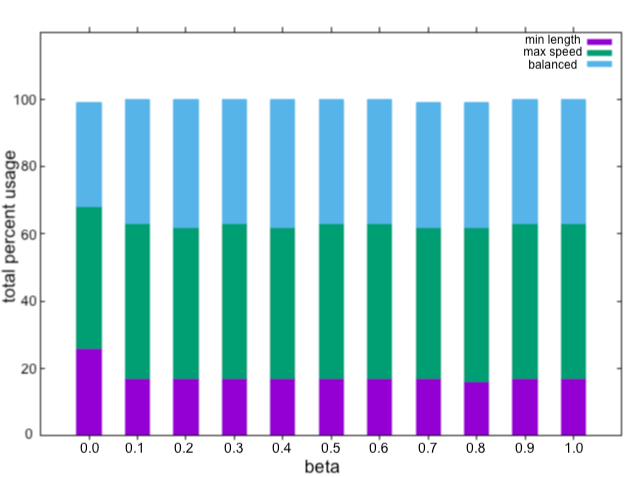}
    \caption{Manhattan}
    \vspace{2em}
    \label{fig:overhead1}
  \end{subfigure}
  \begin{subfigure}{8cm}
    \centering
    \includegraphics[width=8cm]{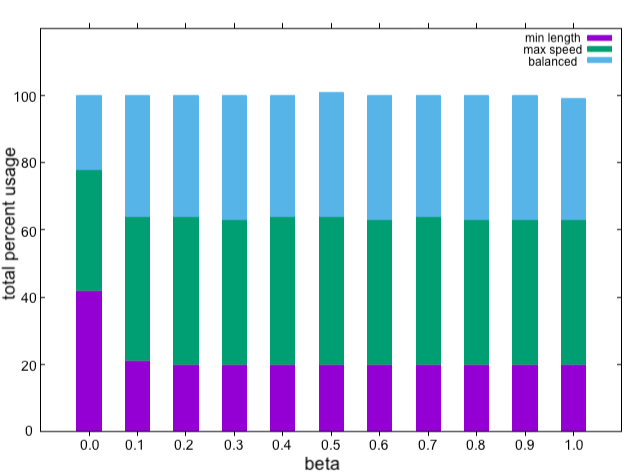}
    \caption{Duluth}
    \vspace{2em}
    \label{fig:overhead2}
  \end{subfigure}
  \begin{subfigure}{8cm}
    \centering
    \includegraphics[width=8cm]{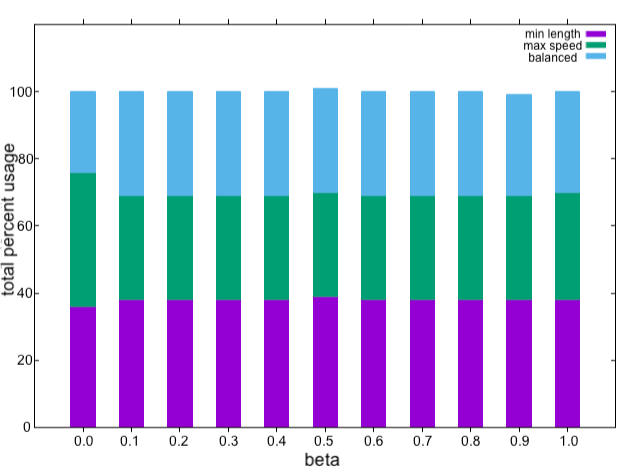}
    \caption{Boulder}
    \label{fig:overhead3}
  \end{subfigure}
  \begin{subfigure}{8cm}
    \centering
    \includegraphics[width=8cm]{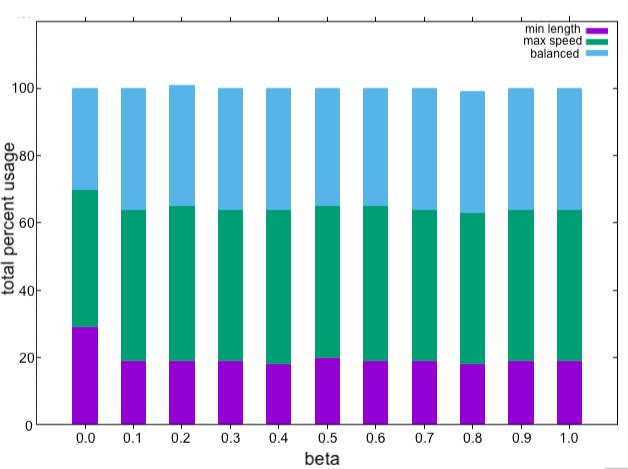}
    \caption{Annapolis}
    \label{fig:overhead4}
  \end{subfigure}
  \caption{Distribution of selected routers that route cars based on minimum length of streets (\textit{minlength}), maximum speed allowed on streets (\textit{maxspeed}), or based on a combination of the two (\textit{balanced}) for different $\beta$ values and different traffic settings.}
  \label{fig:routers}
\end{figure}

We then turned our attention to EPOS plans.  
For the balancing effect to take place, (i) some cars need to select a different router from their preferred one, (ii) the different router needs to provide a different route than the one provided by the preferred router. 
To investigate (i), we plotted the percentage of router utilization for each setting and for each \mybeta value (Figure~\ref{fig:routers}).
In all settings, the distribution of router utilization changed significantly when setting \mybeta to 0.
This means that EPOS did indeed force several cars to select non-preferred routers in all settings. 

These observations show that neither the structural properties of the particular networks nor particular route plans of EPOS explain the insignificant effect on trip overheads in certain cases. Therefore, the clear effect of traffic load on the network of Eichst{\"a}tt is hypothesized as the one explaining the low performance in Boulder and Annapolis. Future work will confirm this using a large-scale computational infrastructure to support such ambitious research.


\section{Conclusions and future work}
\label{sec:conclusions}
In this paper, we focused on new mobility concepts in smart cities and investigated the use of multi-agent learning in optimizing route planning. 
In particular, considering each (potentially autonomous) car as an agent that has several plans, i.e. routes to a destination, and can considers both local and global objectives, we investigated whether increasing the altruism of the agents can have a positive effect on the overall performance of traffic under varying traffic levels. 
We performed detailed measurements to answer the above question using a simulation framework that integrates SUMO, a well-known traffic simulator, and EPOS, a decentralized agent-based framework. 
Our study focused on rush hour traffic in four US cities and found that (i) load balancing can indeed be achieved by increasing the agents' altruism, (ii) whether a positive effect on network performance can be observed depends on the characteristics of the cities and, in particular, on the density of the city network as well as (iii) the level of traffic in the network.

As future work, we would like to compare further cities under varying traffic level and relate their (topological) characteristics with their capacity to optimize traffic flows. Finally, a very interesting direction of research concerns the addition of the fairness objective in the decision making of agents in our setting.

\bibliographystyle{elsarticle-num}
\bibliography{refs}

\begin{thebibliography}{10}
\expandafter\ifx\csname url\endcsname\relax
  \def\url#1{\texttt{#1}}\fi
\expandafter\ifx\csname urlprefix\endcsname\relax\def\urlprefix{URL }\fi
\expandafter\ifx\csname href\endcsname\relax
  \def\href#1#2{#2} \def\path#1{#1}\fi

\bibitem{bell1997transportation}
M.~G.~H. Bell, Y.~Iida, Transportation network analysis, Wiley, 1997.

\bibitem{merchant1978model}
D.~K. Merchant, G.~L. Nemhauser, A model and an algorithm for the dynamic
  traffic assignment problems, Transportation Science 12~(3) (1978) 183--199.

\bibitem{merchant1978optimality}
D.~K. Merchant, G.~L. Nemhauser, Optimality conditions for a dynamic traffic
  assignment model, Transportation Science 12~(3) (1978) 200--207.

\bibitem{peeta2001foundations}
S.~Peeta, A.~K. Ziliaskopoulos, Foundations of dynamic traffic assignment: The
  past, the present and the future, Networks and Spatial Economics 1~(3-4)
  (2001) 233--265.

\bibitem{friesz1989dynamic}
T.~L. Friesz, J.~Luque, R.~L. Tobin, B.-W. Wie, Dynamic network traffic
  assignment considered as a continuous time optimal control problem,
  Operations Research 37~(6) (1989) 893--901.

\bibitem{ran1993new}
B.~Ran, D.~E. Boyce, L.~J. LeBlanc, A new class of instantaneous dynamic
  user-optimal traffic assignment models, Operations Research 41~(1) (1993)
  192--202.

\bibitem{chen1998game}
O.~Chen, M.~Ben-Akiva, Game-theoretic formulations of interaction between
  dynamic traffic control and dynamic traffic assignment, Transportation
  Research Record 1617~(1) (1998) 179--188.

\bibitem{mahmassani1993network}
H.~S. Mahmassani, S.~Peeta, Network performance under system optimal and user
  equilibrium dynamic assignments: implications for ATIS, Transportation
  Research Board, 1993.

\bibitem{jahn2005system}
O.~Jahn, R.~H. M{\"o}hring, A.~S. Schulz, N.~E. Stier-Moses, System-optimal
  routing of traffic flows with user constraints in networks with congestion,
  Operations Research 53~(4) (2005) 600--616.

\bibitem{groot2014toward}
N.~Groot, B.~De~Schutter, H.~Hellendoorn, Toward system-optimal routing in
  traffic networks: A reverse stackelberg game approach, IEEE Transactions on
  Intelligent Transportation Systems 16~(1) (2014) 29--40.

\bibitem{shen2014system}
W.~Shen, H.~M. Zhang, System optimal dynamic traffic assignment: Properties and
  solution procedures in the case of a many-to-one network, Transportation
  Research Part B: Methodological 65 (2014) 1--17.

\bibitem{zhu2015linear}
F.~Zhu, S.~V. Ukkusuri, A linear programming formulation for autonomous
  intersection control within a dynamic traffic assignment and connected
  vehicle environment, Transportation Research Part C: Emerging Technologies 55
  (2015) 363--378.

\bibitem{levin2015intersection}
M.~W. Levin, S.~D. Boyles, Intersection auctions and reservation-based control
  in dynamic traffic assignment, Transportation Research Record 2497~(1) (2015)
  35--44.

\bibitem{levin2016multiclass}
M.~W. Levin, S.~D. Boyles, A multiclass cell transmission model for shared
  human and autonomous vehicle roads, Transportation Research Part C: Emerging
  Technologies 62 (2016) 103--116.

\bibitem{trapp}
I.~Gerostathopoulos, E.~Pournaras,
  \href{https://dl.acm.org/citation.cfm?id=3341532}{{TRAPP}ed in traffic?: a
  self-adaptive framework for decentralized traffic optimization}, in:
  Proceedings of the 14th International Symposium on Software Engineering for
  Adaptive and Self-Managing Systems, SEAMS@ICSE 2019, Montreal, QC, Canada,
  May 25-31, 2019, 2019, pp. 32--38.
\newline\urlprefix\url{https://dl.acm.org/citation.cfm?id=3341532}

\bibitem{kazhamiakin2015using}
R.~Kazhamiakin, A.~Marconi, M.~Perillo, M.~Pistore, G.~Valetto, L.~Piras,
  F.~Avesani, N.~Perri, Using gamification to incentivize sustainable urban
  mobility, in: 2015 IEEE First International Smart Cities Conference (ISC2),
  IEEE, 2015, pp. 1--6.

\bibitem{klein2018emergence}
I.~Klein, E.~Ben-Elia, Emergence of cooperative route-choice: A model and
  experiment of compliance with system-optimal atis, Transportation Research
  Part F: Traffic Psychology and Behaviour 59 (2018) 348--364.

\bibitem{ringhand2018make}
M.~Ringhand, M.~Vollrath, Make this detour and be unselfish! influencing urban
  route choice by explaining traffic management, Transportation Research Part
  F: Traffic Psychology and Behaviour 53 (2018) 99--116.

\bibitem{lopez_microscopic_2018}
P.~A. Lopez, M.~Behrisch, L.~Bieker-Walz, J.~Erdmann, Y.~Flötteröd,
  R.~Hilbrich, L.~Lücken, J.~Rummel, P.~Wagner, E.~WieBner, Microscopic
  {Traffic} {Simulation} using {SUMO}, in: 2018 21st {International}
  {Conference} on {Intelligent} {Transportation} {Systems} ({ITSC}), 2018, pp.
  2575--2582.
\newblock \href {http://dx.doi.org/10.1109/ITSC.2018.8569938}
  {\path{doi:10.1109/ITSC.2018.8569938}}.

\bibitem{7906642}
L.~{Codeca}, R.~{Frank}, S.~{Faye}, T.~{Engel}, Luxembourg sumo traffic (lust)
  scenario: Traffic demand evaluation, IEEE Intelligent Transportation Systems
  Magazine 9~(2) (2017) 52--63.
\newblock \href {http://dx.doi.org/10.1109/MITS.2017.2666585}
  {\path{doi:10.1109/MITS.2017.2666585}}.

\bibitem{8275627}
L.~{Codecá}, J.~{Härri}, Towards multimodal mobility simulation of c-its: The
  monaco sumo traffic scenario, in: 2017 IEEE Vehicular Networking Conference
  (VNC), 2017, pp. 97--100.
\newblock \href {http://dx.doi.org/10.1109/VNC.2017.8275627}
  {\path{doi:10.1109/VNC.2017.8275627}}.

\bibitem{Wegener:2008:TIC:1400713.1400740}
A.~Wegener, M.~Pi\'{o}rkowski, M.~Raya, H.~Hellbr\"{u}ck, S.~Fischer, J.-P.
  Hubaux, \href{http://doi.acm.org/10.1145/1400713.1400740}{{TraCI}: An
  interface for coupling road traffic and network simulators}, in: Proceedings
  of the 11th Communications and Networking Simulation Symposium, CNS '08, ACM,
  New York, NY, USA, 2008, pp. 155--163.
\newblock \href {http://dx.doi.org/10.1145/1400713.1400740}
  {\path{doi:10.1145/1400713.1400740}}.
\newline\urlprefix\url{http://doi.acm.org/10.1145/1400713.1400740}

\bibitem{epos2018}
E.~Pournaras, P.~Pilgerstorfer, T.~Asikis,
  \href{http://doi.acm.org/10.1145/3277668}{Decentralized collective learning
  for self-managed sharing economies}, ACM Trans. Auton. Adapt. Syst. 13~(2)
  (2018) 10:1--10:33.
\newblock \href {http://dx.doi.org/10.1145/3277668}
  {\path{doi:10.1145/3277668}}.
\newline\urlprefix\url{http://doi.acm.org/10.1145/3277668}

\end{thebibliography}

\end{document}